\documentclass[conference]{IEEEtran}
\IEEEoverridecommandlockouts

\usepackage{amsmath}  % mathematics
\usepackage{amssymb}
\usepackage{mathrsfs}

\usepackage{balance}  % two-column mode

\usepackage{cite}
\usepackage{url}

\usepackage{ifpdf} % conditional compilation - pdf or dvi

\usepackage{multicol}  % misc
\usepackage{makeidx}
\usepackage{color}
\usepackage{epsf}
\usepackage{subfigure}
\usepackage{comment}

\includecomment{nonDraftFigure}
\excludecomment{draftFigure}

\makeatletter
\def\normalsize{\@setfontsize{\normalsize}{10bp}{10.00pt}}
\normalsize
\makeatother

\ifCLASSINFOpdf
   \usepackage[pdftex]{graphicx}
   \graphicspath{{../fig/pdf/}}
   \DeclareGraphicsExtensions{{.pdf}}
\else
   \usepackage[dvips]{graphicx}
   \graphicspath{{../fig/eps/}}
   \DeclareGraphicsExtensions{{.eps}}
\fi

\def\refname{\normalsize \centerline{\bf REFERENCES} }

\begin{document}

\title{An Iterative Noncoherent Relay Receiver \\ for the Two-way Relay Channel}

\author{
Terry Ferrett\IEEEauthorrefmark{1}, 
Matthew C. Valenti\IEEEauthorrefmark{1}
Don Torrieri\IEEEauthorrefmark{2}\\

\IEEEauthorrefmark{1}West Virginia University, Morgantown, WV, USA.\\
\IEEEauthorrefmark{2}U.S. Army Research Laboratory, Adelphi, MD, USA.
\vspace{-0.35cm}
\thanks{M.C. Valenti and T. Ferrett were sponsored by the National Science Foundation under Award No. CNS-0750821.}
}

\maketitle

\begin{abstract}
Physical-layer network coding improves the throughput of the two-way relay channel by allowing multiple source terminals to transmit simultaneously to the relay.  However, it is generally not feasible to align the phases of the multiple received signals at the relay, which motivates the exploration of noncoherent solutions.  In this paper, turbo-coded orthogonal multi-tone frequency-shift keying (FSK) is considered for the two-way relay channel. In contrast with {\em analog} network coding, the system considered is an instance of {\em digital} network coding; i.e., the relay decodes the network codeword and forwards a re-encoded version.  Crucial to noncoherent digital network coding is the implementation of the relay receiver, which is the primary focus of the paper.    The relay receiver derived in this paper supports any modulation order that is a power of two, and features the iterative feedback of a priori information from the turbo channel decoder to the demodulator; i.e., it uses \emph{bit interleaved coded modulation with iterative decoding} (BICM-ID). The performance of the receiver is investigated in Rayeligh fading channels through error-rate simulations and a capacity analysis.  Results show that the BICM-ID receiver improves energy efficiency by $0.5$-$0.9$ dB compared to a non-iterative receiver implementation.
\end{abstract}

\IEEEpeerreviewmaketitle

\section{Introduction}
In a two-way relay channel (TWRC), a pair of \emph{terminal nodes} exchange information using a \emph{relay node}. Each terminal is both a source and a destination. The terminals are assumed to have no direct radio link.
\emph{Physical-layer network coding} (PNC) \cite{zhang:2006} is a technique that may be applied in the TWRC to increase throughput over conventional techniques by eliminating the number of transmissions necessary for information exchange between the terminals.
The exchange is broken into two phases: the \emph{multiple-access} (MA) phase and the \emph{broadcast} phase.
In the MA phase, the terminals transmit simultaneously to the relay, which receives the electromagnetic sum of the transmitted signals.
In the broadcast phase, the relay broadcasts the sum of information to the terminals, each of which detects the other terminal's information by subtracting its own signal from the received sum.

PNC may be implemented using one of two schemes: \emph{analog network coding} (ANC) and \emph{digital network coding} (DNC).  In ANC, the relay forwards the received signal sum directly and all of the processing is performed at the terminals.  While the benefit of ANC is a simple relay implementation, the disadvantage is that the noise at the relay is also forwarded to the terminals and the processing requirements at the terminals can be burdensome.  In DNC,  the relay performs detection of the network-coded bits, essentially cleaning up the noise at the relay.  It then remodulates the signal and broadcasts to the terminals.  The benefit of DNC is that the noise received at the relay is not retransmitted and the terminal receivers are simplified, but the disadvantage is that a more complex receiver is required at the relay.  Thus, a crucial aspect of implementing PNC is the formulation of an efficient relay receiver, and the selection of coded-modulation formats that work well with the DNC system.  

In \cite{vtf:2011}, we found that turbo-coded binary frequency-shift keyed (FSK) modulation is a good candidate for DNC, and formulated a noncoherent receiver.
Noncoherent FSK is useful in scenarios where the received signal phase is corrupted and cannot be tracked accurately.
Examples of phase corrupting effects include imperfect or unsynchronized oscillators and Doppler shift.
Tracking phase is even more challenging for a DNC relay receiver than for a conventional point-to-point receiver, as the relay must track two phases simultaneously rather than one.
Noncoherently detecting the FSK signal eliminates the phase tracking requirement.
Multi-tone orthogonal FSK modulation is beneficial when energy efficiency has higher priority than bandwidth efficiency.
If $\mathcal{E}_b/N_0$ is held constant, then increasing the modulation order increases the energy per symbol,
which increases the minimum distance between symbols, leading to lower error rates.
However, bandwidth usage is proportional to the modulation order, implying a trade-off between
bandwidth and energy efficiency.

%The network-coding operation is defined as the modulo-two sum of the binary information labeling the symbols transmitted by the sources.
%The goal of the relay receiver is to detect the network-coded combination of information bits transmitted by the source nodes.

A particular technique for combining binary channel coding and $M$-ary modulation with $M > 2$ is \emph{bit-interleaved coded modulation} (BICM) \cite{caire:1998}.
A binary channel codeword is generated, interleaved, and passed to an $M$-ary modulator, which maps codeword bits to symbols
for channel transmission.
The receiver demodulates the symbols, producing soft estimates of each bit.
A binary soft-input channel decoding scheme is applied to the soft estimates.
The energy efficiency of BICM can be improved by feeding back information from the channel decoder to the demodulator,
and performing iterative detection.
The notion of feeding back information from decoder to demodulator is called \emph{BICM with iterative decoding} (BICM-ID) \cite{li:1997}.

The present work extends \cite{vtf:2011} by considering turbo-coded multi-tone FSK modulation, and deriving an iterative receiver capable of performing iterative $M-ary$ demodulation and turbo-decoding. 
The key contribution of this work is a DNC relay receiver capable of noncoherent operation which utilizes BICM-ID for improved performance.
BICM-ID improves the error rate performance of the relay receiver versus BICM, while adding computational complexity to the receiver, introducing a trade-off in receiver design. 
A receiver formulation considering BICM-ID and FSK modulation for single-source, point-to-point channels is given by \cite{valenti:2005}.
The present work extends the receiver in \cite{valenti:2005} to support the two-source TWRC model and digital network coding.

A Turbo BICM channel decoding scheme for the relay in the MA phase of DNC is given by \cite{zhan:2010}.
In contrast to the current work, the scheme in \cite{zhan:2010} considers only coherent reception
and two-dimensional modulation.
Further, this work does not consider the BICM-ID feedback scheme from decoder to demodulator.
\cite{wu:2011} gives a channel model for the DNC MA phase that considers symbol and frame asynchrony between transmissions
by the terminals.
Techniques for optimal detection of network-coded information at the relay employing LDPC channel coding and BPSK modulation under symbol and frame misalignment are presented.
To our knowledge, no prior work has considered the application of BICM-ID to a noncoherent DNC relay receiver.

In general, there are several ways to implement channel-coded DNC depending on the specific order in which demodulation, channel and network coding are applied at the relay.
In \cite{zhang:2009}, the performance of different channel-coded DNC techniques are compared.
The technique in the present work, in which the relay demodulator forms likelihood ratios of the network-coded bits for channel decoding, is contrasted with a technique in which the demodulator performs decoding using probability mass functions of the arithmetic sum of received symbols at the relay.
It is shown that the technique utilizing arithmetic sums achieves superior capacity to bit-level network coding scheme in particular SNR regions, however, the specific channel code applied to arithmetic network coding is a repeat-accumulate code designed specifically for the scheme, and the extension to codes such as LDPC or Turbo requires modifications to existing channel coding algorithms, which have not yet been considered.
Our usage of the network-coded bit scheme is based on the assumption that its implementation in existing systems requires no modification to existing channel coding schemes, which is a practical advantage.

An outline of the remainder of this paper is as follows.
Section \ref{sec:sysm} develops the system model used throughout the paper.
Section \ref{sec:demod} presents the development of the relay demodulator capable of performing digital-network coded BICM-ID at the relay in the TWRC.
Section \ref{sec:sim} presents simulated error rate performance and capacity analysis of the relay receiver.

\section{System Model}\label{sec:sysm}
\begin{figure}[b]
\centering
\includegraphics[width=9cm]{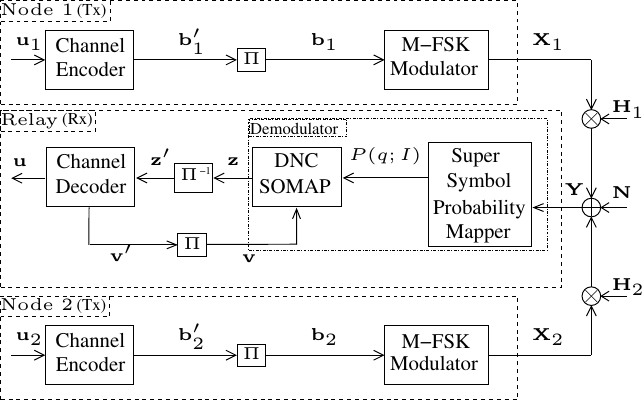}
\caption{System Model - TWRC DNC MAC Phase}
\label{fig:sysm}
\end{figure}

This section presents the system model used throughout the work.
Modulation, channel coding, channel model, and relay reception are described in precise detail.
An overview of the iterative decoding process at the relay is given.
The system model illustrating transmission by the terminal nodes to the relay is shown in Fig. \ref{fig:sysm}.

\subsection{Transmission by Terminal Nodes}

The \emph{terminal} nodes $\mathcal{N}_i, \ i \in \{1,2\}$ generate binary information sequences $\mathbf{u}_i = [u_{1,i}, ..., u_{K,i}]$ having length $K$. 
A rate-$r_S$ turbo code is applied to each $\mathbf{u}_i$, generating a length $L = K/r_S$ binary channel codeword, denoted by $\mathbf{b}'_i = [b_{1,i}, ..., b_{L,i}]$.
The codeword is passed through an interleaver, modeled as a permutation matrix $\mathbf{\Pi}$ having dimensionality $L \times L$: $\mathbf{b}_i =  \mathbf{b}'_i \mathbf{\Pi}$. 
Let $\mathcal{D} = \{ 0,...,M-1\}$ denote the set of integer indices corresponding to each FSK tone, where $M$ is the modulation order.
 The number of bits per symbol is $\mu = \log_2M$.
 The codewords $\mathbf{b}_i$ at each node are divided into $N_q = L/\mu$ sets of bits, which are passed to an $M$-ary FSK modulator.
 The modulator maps each set to an $M$-ary symbol $q_{k,i} \in \mathcal{D}$, where $k$ denotes the symbol number, and $i$ denotes the terminal.
The modulated signal transmitted by terminal $\mathcal{N}_i$ during signaling interval $kT_s \leq t \leq (k+1)T_s$ is
\vspace{-1mm}
\begin{eqnarray}\label{eqn:TxSignal}
s_{k,i}(t) =  \sqrt{ \frac{2}{T_s} }
\cos
\left[
  2 \pi
  \left(
    f_{i}
    + \frac{q_{k,i}}{T_s}
    \right )
    (t - kT_s)
    \right]
\end{eqnarray}

\noindent where $s_{k,i}(t)$ is the transmitted signal, $f_{i}$ is the carrier frequency of terminal $\mathcal{N}_i$, and $T_s$ is the symbol period. 

 The continuous-time signals $s_{k,i}(t)$ are represented in discrete time by the set of column vectors $\{\mathbf{e}_{q_{k,i}}: q_{k,i} \in \mathcal{D} \}$.
  The column vector $\mathbf{e}_{q_{k,i}}$ is length $M$, contains a 1 at vector position $q_{k,i}$, and $0$ elsewhere.
  The modulated codeword from terminal $\mathcal{N}_i$ is represented by the matrix of symbols $\mathbf{X}_i =  [\mathbf{x}_{1,i}, ..., \mathbf{x}_{N_q,i}]$, having dimensionality $M \times N_q$, where $\mathbf{x}_{k,i} = \mathbf{e}_{q_{k,i}}$.

\noindent

\subsection{Channel Model}

All channels are modeled as flat-fading channels having independent gains for every symbol period.
The complex-valued channel gain from node $\mathcal{N}_i$ to the relay during a particular signaling interval $k$ is denoted by $h_{k,i}$.
The gain is represented as $h_{k,i} = \alpha_{k,i} e^{j \theta_{k,i}}$, where $\alpha_{k,i}$ is the received amplitude and $\theta_{k,i}$ is the phase, which depends on the phase shift of the channel and corruption in the reference signals at the terminal and relay induced by hardware imperfections.
The amplitudes of the gains are selected such that the received energy at the relay from node $\mathcal{N}_i$ is $\mathcal{E}_i$
\vspace{-5mm}

\begin{eqnarray}\label{eqn:fading_amplitude_energy}
\mathcal{E}_i & = & E[|h_{k,i}|^2] = E[\alpha^2_{k,i}].
\end{eqnarray}

Consider transmission of a single frame of $N_q$ symbols to the relay.
The received frame is

\vspace{-5mm}
\begin{align} \label{eqn:rec_sym}
\mathbf{Y} = \mathbf{X}_1 \mathbf{H}_1 + \mathbf{X}_2 \mathbf{H}_2 + \mathbf{N}
\end{align}

\noindent where $\mathbf{H}_i$ is an $N_q \times N_q$ diagonal matrix of channel coefficients having value $h_{k,i}$ at matrix entry $(n,n)$ and $0$ elsewhere, and $\mathbf{N}$ is an $M \times N_q$ noise matrix.
A single pair of channel-corrupted symbols received at the relay is denoted by $\mathbf{y}$, and defined as a \emph{channel observation}. 
In terms of this definition, $\mathbf{Y} =  [\mathbf{y}_{1}, ..., \mathbf{y}_{N_q}]$, where $\mathbf{y}_k$ denotes the $k$-th channel observation. 
Denote the $k$-th column of $\mathbf{N}$ by $\mathbf{n}_k$.
Each column is composed of zero-mean circularly symmetric complex Gaussian random variables having covariance matrix $N_0 \mathbf{I}_M$; i.e., $\mathbf{n}_k \sim \mathcal{N}_c(\mathbf{0}, N_0 \mathbf{I}_M)$.
$N_0$ is the one-sided noise spectral density, and $\mathbf{I}_M$ is the $M$-by-$M$ identity matrix.

\subsection{Relay Reception}

The goal of the relay receiver is to detect the network-coded combination of information bits transmitted by the terminals, $\mathbf{u} = \mathbf{u}_1 \oplus \mathbf{u}_2$.
The relay receiver takes as input the frame of channel observations $\mathbf{Y}$. 
The symbols transmitted by the terminals are assumed perfectly synchronized at the receiver.
Demodulation and iterative channel-decoding are applied to the received frame to detect $\mathbf{u}$.
Define the network codeword as
\vspace{-5mm}

\begin{align}
\mathbf{b} =& \mathbf{b}_1 \oplus \mathbf{b}_2 \nonumber \\
           =& [\ b_{1,1} \oplus b_{1,2} \  ... \ b_{1,L} \oplus b_{2,L} \ ]
\end{align}
\vspace{-3mm}

\noindent Since the turbo code is a systematic linear code, $\mathbf{b}$ forms a code from the codebooks used by the terminal nodes, thus, the channel decoding operation yields a hard decision on the network coded message bits $\mathbf{u}$.

 The demodulator and decoder are implemented using the soft-input soft-output algorithm described in \cite{benedetto:1998}.
  The demodulator takes as input the matrix of received symbols $\mathbf{Y}$ and a-priori probability (APP) information $\mathbf{v}$ and produces extrinsic information $\mathbf{z}$, $\mathbf{v} = [ v_1,..., v_L ]$, $\mathbf{z} = [ z_1, ..., z_L ]$.
Precise description of the quantities $v_m$ and $z_m$ is provided in Section \ref{sec:demod}. 
The extrinsic information is deinterleaved to produce $\mathbf{z}' = \mathbf{z} \mathbf{\Pi}^{-1}$ and passed to the decoder.
  The decoder refines the estimate of $\mathbf{z}'$, producing new extrinsic information $\mathbf{v}'$ which is interleaved to produce $\mathbf{v} = \mathbf{v}' \mathbf{\Pi}$ and returned to the demodulator. 

The average symbol signal-to-noise ratio transmitted by each terminal $\mathcal{E}_i/N_0 \ i \in \{1,2\}$ is known at the demodulator.
The demodulator may operate using several cases of knowledge of the channel gains $h_{k,i}$ (termed channel state information or CSI): the case in which the gains are completely known,
the case in which only the fading amplitudes $\alpha_{k,i}$ are known, and the case in which no information about the gain is known (no CSI).

Specific details of the demodulator are discussed in Section \ref{sec:demod}.
The details of the channel decoder have been treated extensively in the literature \cite{benedetto:1998}, \cite{robertson:1997}, and will not be discussed here.
Note that a feed-forward BICM receiver without BICM-ID does not feed back extrinsic information from the decoder to the demodulator.

The relay encodes and modulates the network-decoded message bits $\mathbf{u}$ and broadcasts its modulated signal to the terminal nodes. 
The signal traverses two independent fading channels, and the terminal nodes receive independently faded versions of the message bits corrupted by white noise.
The terminal nodes demodulate and decode the signal received from the relay to form estimates of $\mathbf{u}$.
Let $\hat{\mathbf{u}}$ denote the detected sequence at $\mathcal{N}_1$ and $\tilde{\mathbf{u}}$ denote the detected sequence at $\mathcal{N}_2$.
Each terminal estimates the information bits transmitted by the opposite node by subtracting its own information sequence from
the sequence detected from transmission by the relay: $\mathbf{g}_2 = \hat{\mathbf{u}} \oplus \mathbf{u}_1$ at $\mathcal{N}_1$ and
$\mathbf{g}_1 = \tilde{\mathbf{u}} \oplus \mathbf{u}_2$ at $\mathcal{N}_2$.
Since the links from the relay to the terminals are conventional point-to-point links with no interfering transmissions, specific details of the terminal-node receivers are omitted.

\section{Soft N-FSK Relay Demodulator}\label{sec:demod}
The soft-output relay demodulator maps the received sum of symbols from the terminals to log-likelihood ratios of the network-coded bits.
 The demodulator operates iteratively, using extrinsic information fed back from the channel decoder to refine the information prior to each decoding iteration.
After a specified number of iterations has been reached, the decoder makes a hard decision on the network-coded bits.

The demodulator processes a frame of channel observations $\mathbf{Y}$ one observation at a time.
Since the operation performed on each observation is the same, we may drop the dependence on a particular
signaling interval in the frame to simplify the notation.
Denote a single received channel observation as $\mathbf{y}$.
During the first demodulation and decoding iteration, the demodulator computes the probability of every possible combination of symbols transmitted by the terminals: $P(q;I)$, where $q$ is defined as the tuple
\vspace{-5mm}

\begin{align}
q = (q_1,q_2) \ \ \ q_1,q_2 \in \mathcal{D} \ \ \ q \in \mathcal{D} \times \mathcal{D}
\end{align}

\noindent and $q_1$ and $q_2$ denote the symbols from terminal $\mathcal{N}_1$
and $\mathcal{N}_2$, respectively.
We will refer to $q$ as a \emph{super-symbol}.
Define this operation as the \emph{super-symbol probability mapping} stage.
The symbol probabilities $P(q;I)$ are fixed for all demodulation and decoding iterations.
Since the cardinality of $\mathcal{D} \times \mathcal{D}$ is $M^2$, the  relay receiver computes $M^2$ probabilities, versus a conventional point-to-point receiver which only computes $M$ probabilities, as only one terminal is present.

On the first and subsequent decoding iteration, the symbol probabilities are transformed to the set of $\mu$ log-likelihood ratios associated with each network-coded bit mapped to the super-symbol.
 Denote this operation as \emph{digital network-coded soft mapping} (DNC-SOMAP).
A general description of SOMAP for the point-to-point channel is given by \cite{benedetto:1998}.
The DNC-SOMAP takes as input the symbol probabilities $P(q;I)$ and extrinsic information represented by bit probabilities associated with each network coded bit $P(\mathbf{c};I)$ fed back from the channel decoder, where $\mathbf{c}$ denotes the $\mu$ network coded bits mapped to $q$, $P(\mathbf{c};I) = \{P(c_k;I), 0 \leq k \leq \mu -1 \}$, $c_k = c_{1,k} \oplus c_{2,k}$, and $c_{i,k}$ denotes the $k$-th bit mapped to symbol $q_i$. $i \in \{1,2\}$.
On the first demodulation iteration, no decoding has been performed, and the bit probabilities are assumed equally likely.
The DNC-SOMAP produces estimated probabilities of values taken by $\mathbf{c}$: $P(\mathbf{c};O) = \{P(c_k;I), 0 \leq k \leq \mu -1 \}$.

The input distributions to the DNC-SOMAP with respect to the super-symbol are represented as probabilities,
while the input and output distributions with respect to the network-coded bits mapped to each super-symbol are represented as log-likelihood ratios.
Log-likelihood representation facilitates soft-decision decoding.
The input representing the log-likelihood ratio of the $m$-th bit mapped to the super-symbol is related to the input distribution by

\vspace{-2mm}
\begin{align}\label{eq:input_bit}
v_k = \log \frac{ P(c_k = 1; I) }{ P(c_k = 0; I) }, \ 0 \leq k \leq \mu - 1 .
\end{align}

\noindent The output representing the log-likelihood ratio of the $k$-th bit mapped to the super-symbol is related to the output distribution by

\vspace{-2mm}
\begin{align} \label{eq:somap_out_llr}
z_k = \log \frac{ P(c_k = 1; O) }{ P(c_k = 0; O) }, \ 0 \leq k \leq \mu - 1 .
\end{align}

The DNC-SOMAP output distribution is related to the input distributions by
\vspace{-2mm}
\begin{align}\label{eq:somap_out_distribution_symbolic}
P(c_k=\ell; O) = \sum_{\begin{subarray}  (q: c_k = \ell \end{subarray}}  p(\mathbf{y} |q) \prod_{\begin{subarray} jj=0 \\j \neq m \end{subarray}}^{\mu-1} P( c_j ; I)
\end{align}

\noindent Substituting the specific values of the distribution (\ref{eq:input_bit}) into the expression for output (\ref{eq:somap_out_distribution_symbolic}),

\vspace{-2mm}
\begin{align}\label{eq:somap_out_distribution_specific}
P( c_k = \ell; O) =  \sum_{\begin{subarray} (q: c_k = \ell \end{subarray}}  p(\mathbf{y} |q)  \prod_{\begin{subarray} jj=0 \\j \neq m \end{subarray}}^{\mu-1}  \frac{ e^{ c_j v_j} }{ 1+e^{v_j} } 
\end{align}

The output log-likelihood ratio of the DNC-SOMAP may be found by combining (\ref{eq:somap_out_distribution_specific}) and (\ref{eq:somap_out_llr}):

\vspace{-2mm}
\begin{align}\label{eq:somap_out_llr_full}
z_k = \log \frac{ \displaystyle\sum_{\begin{subarray} (q: c_k = 1 \end{subarray}} p(\mathbf{y} |q)  \prod_{\begin{subarray} jj=0 \\j \neq m \end{subarray}}^{\mu-1}   e^{ c_j  v_j}  }
 { \displaystyle\sum_{\begin{subarray} (q: c_k = 0 \end{subarray}} p(\mathbf{y} |q)  \prod_{\begin{subarray} jj=0 \\j \neq m \end{subarray}}^{\mu-1}    e^{c_j v_j}  }
\end{align}
\noindent where the term $(1+e^{v_j})$ cancels in the ratio.
For the purpose of numeric implementation, it is useful to simplify this expression using the \emph{max-star} operator

\vspace{-2mm}
\begin{align}
\underset{i}{\operatorname{max} \hspace{-0.5mm}*} \{ x_i \} = \log \left\{ \sum_i e^{ x_i } \right\}
\end{align}

\noindent where the binary max-star operator is  $\max*(x,y) = \max(x,y) + \log( 1 + e^{ -|x-y| } ) $ and
multiple arguments imply a recursive relationship; for example: $\max*(x,y,z) = \max*( x, \max*(y,z) )$.
Applying the max-star operator to (\ref{eq:somap_out_llr_full})

\vspace{-2mm}
\begin{align} \label{eq:somap_out_llr_maxstar}
 z_k & =  \underset{\begin{subarray} (q: c_k = 1 \end{subarray}}{\operatorname{max}  \hspace{-0.5mm} *} \left[ \log p(\mathbf{y} | q) + \sum_{\begin{subarray} jj=0 \\ j \neq k\end{subarray}}^{\mu-1} c_j v_j\right] \nonumber \\ & -\underset{\begin{subarray} (q: c_k = 0 \end{subarray}}{\operatorname{max}  \hspace{-0.5mm} *} \left[ \log p(\mathbf{y} | q) + \sum_{\begin{subarray} jj=0 \\ j \neq k\end{subarray}}^{\mu-1} c_j v_j  \right].
\end{align}

\noindent A non-iterative BICM demodulator does not utilize feedback from the decoder, so it is implemented using (\ref{eq:somap_out_llr_maxstar}) setting all $v_j = 0$.
The values taken by the pdf $p(\mathbf{y}|q)$ are dependent on the available channel state information.
Description of these pdfs is given in the following subsections.

\subsection{Noncoherent Reception with CSI}\label{subsec:csi}

Considering noncoherent reception with CSI, the phases of $h_1$ and $h_2$ are not known.
The pdf of the super-symbol $p(\mathbf{y}|q)$ takes different forms depending on the values of the symbols $q_1$ and $q_2$.
When $q_1$ and $q_2$ are different, the pdf takes the form \cite{vtf:2011}

\vspace{-2mm}
\begin{align}\label{eqn:csi_different}
p(\mathbf{y}|q) =  \exp \left\{ - \frac{\alpha_{1}^2 + \alpha_{2}^2}{N_0} \right\}I_0 \left( \frac{2|y_{q_1}| \alpha_1}{N_0}  \right) I_0 \left( \frac{2|y_{q_2}| \alpha_2}{N_0}  \right)
\end{align}

\noindent where $I_0$ denotes the zeroth-order modified Bessel function of the first kind, and $y_{q_1}$ and $y_{q_2}$ denote the $q_1$ and $q_2$-th entries of the received channel observation $\mathbf{y}$.
When $q_1$ and $q_2$ are the same, the form of the pdf is \cite{vtf:2011}

\vspace{-2mm}
\begin{align}\label{eqn:csi_same}
p(\mathbf{y}|q) =  \exp \left\{ - \frac{\alpha^2}{N_0} \right\} I_0 \left( \frac{2|y_{q_1}| \alpha}{N_0}  \right)
\end{align}

\noindent where $\alpha$ is approximated as $\alpha = \sqrt{ \alpha_1^2 + \alpha_2^2}$.
A discussion of this approximation is found in \cite{vtf:2011}.
The symbol probability mapper computes the values of (\ref{eqn:csi_different}) and (\ref{eqn:csi_same}) which are substituted into (\ref{eq:somap_out_llr_maxstar}) to compute the output log-likelihood ratio of the DNC-SOMAP.

\subsection{Noncoherent Reception Without CSI}\label{subsec:nocsi}

When the relay possesses no knowledge of the channel gains, the pdf $p(\mathbf{y}|q)$ is a function of the average symbol energy transmitted by the terminals and the noise variance.
In the case that the symbols $q_1$ and $q_2$ are different, the pdf takes the form \cite{fvt:2011}

\vspace{-2mm}
\begin{multline}\label{nocsi_diff}
p(\mathbf{y}|q ) = \\
\left[ \left( \frac{1}{\mathcal{E}_1 \mathcal{E}_2} \right) \left( \frac{1}{\mathcal{E}_1} + \frac{1}{N_o}\right) \left( \frac{1}{\mathcal{E}_2} + \frac{1}{N_0} \right) \right]^{-1} \\
\times \exp \left\{ \frac{|y_{q_1}|^2 \mathcal{E}_1}{N_0(N_0 + \mathcal{E}_1)} + \frac{ |y_{q_2}|^2 \mathcal{E}_2}{N_0 (N_0 + \mathcal{E}_2)} \right\}.
\end{multline}

\noindent and when $q_1$ and $q_2$ are the same \cite{fvt:2011}

\vspace{-2mm}
\begin{multline}\label{nocsi_same}
p(\mathbf{y}|q)
= 
 \left( \frac{1}{\mathcal{E}_1 + \mathcal{E}_2}\right) \left( \frac{1}{\mathcal{E}_1 + \mathcal{E}_2} + \frac{1}{N_0} \right)^{-1} \\ 
\times \exp \left\{ \frac{|y_{q_1}|^2 (\mathcal{E}_1 + \mathcal{E}_2)}{N_0^2 + N_0 (\mathcal{E}_1 + \mathcal{E}_2)} \right\}.
\end{multline}

The symbol probability mapper computes the values of (\ref{nocsi_diff}) and (\ref{nocsi_same}) which are substituted into (\ref{eq:somap_out_llr_maxstar}) to compute the output log-likelihood ratio of the DNC-SOMAP.

\section{Simulation Study}\label{sec:sim}
In this section, simulated error rate performance and capacity analysis is shown considering the demodulator derived in Section \ref{sec:demod} under the channel model described in Section \ref{sec:sysm}.
Error rate and capacity is simulated for the MAC phase of DNC two-way relaying only, as the the broadcast phase is equivalent to a pair of point-to-point links, which has been thoroughly discussed in the literature.

Error rate performance is simulated as a function of FSK modulation order, channel state information at the relay, channel decoding iterations, and presence or absence of information feedback from decoder to demodulator via BICM-ID, and signal-to-noise ratio.
All error rate simulations utilize soft-decision channel coding.
The specific channel code is the Turbo code described by the UMTS standard \cite{umts:2006}.
Channel capacity is simulated as a function of channel state information and modulation order.
The results of simulation are interpreted to provide recommendations for relay receiver configuration.

\subsection{Error Rate Performance}

\begin{figure}[t]
\centering
\includegraphics[width=8.8cm]{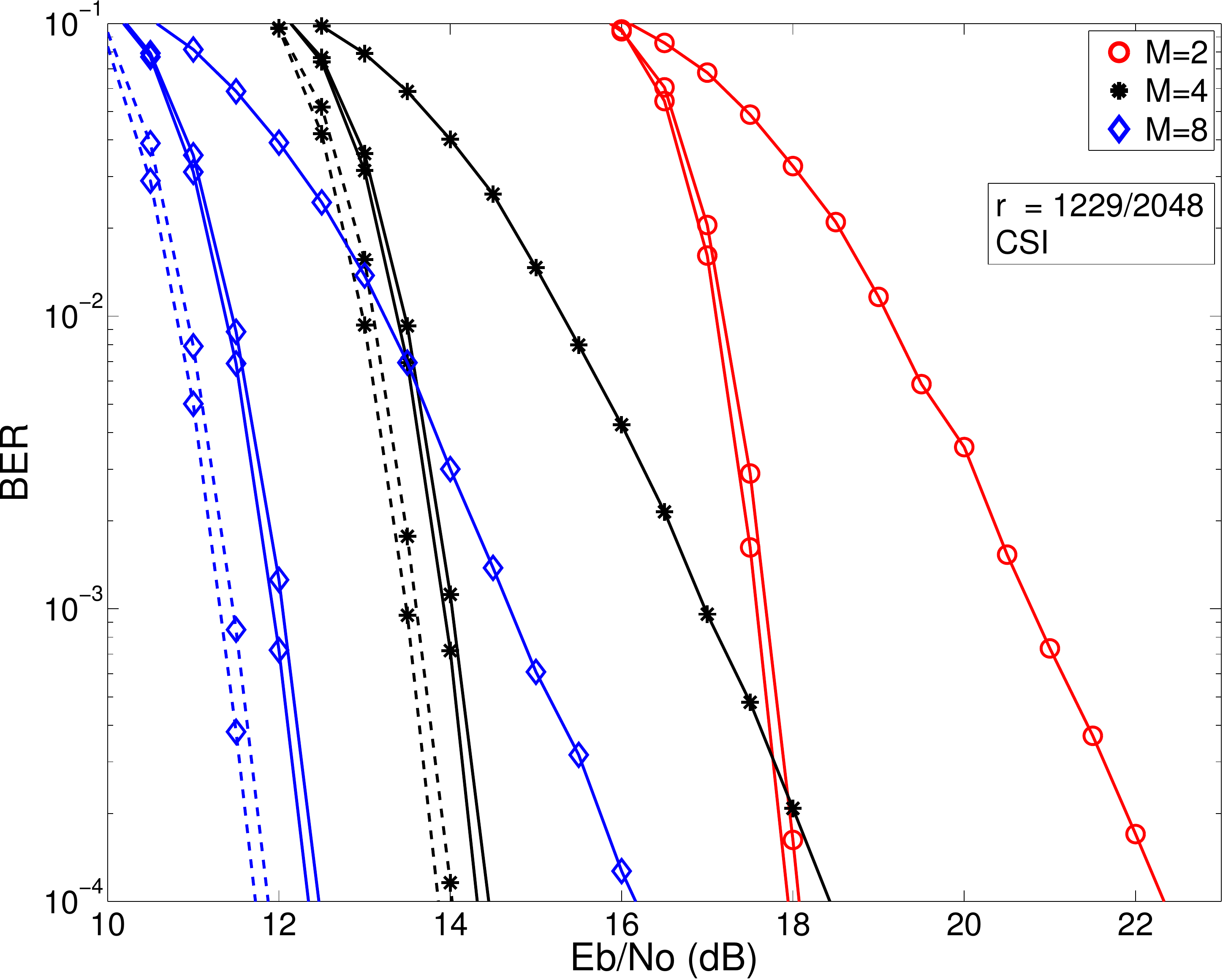}
\caption{
Error-rate performance of the relay demodulator as a function of modulation order and decoding iterations.
The relay receiver possesses CSI.
The data sequence is length K = 1229.
The channel code rate is $r=1229/2048 \approx 0.6$.
Solid lines denote BICM, and dashed lines denote BICM-ID.
With each modulation order, from right to left, the number of decoding iterations is 1, 10, and 30.
}
\label{fig:1}
\end{figure}

In this subsection, the specific results of error rate simulation are presented.
All simulations utilize a UMTS Turbo code with rate $K/L = 1229/2048 \approx 0.6$.
The FSK modulation orders utilized at the terminals are $M = \{2,4,8\}$.
The demodulator is simulated with and without channel state information, as described in
subsections \ref{subsec:csi} and \ref{subsec:nocsi}.
The number of decoding iterations is chosen from $\{1,2,4,30\}$.
The range of decoding iterations was chosen between $1-30$ as negligible performance improvement was observed
beyond this range for the selected system and channel parameters.
Decoding is performed with BICM and BICM-ID for modulation orders $M > 2$.

Error rate performance of the relay demodulator as a function of modulation order and decoding iterations
is shown in Fig. \ref{fig:1}.
Comparing curves utilizing the same number of decoding iterations, increasing $M$ from $2$ to $4$ and $4$ to $8$ improves energy efficiency by $4$ and $2$ dB respectively, regardless of the number of decoding iterations or decoder feedback.
The performance of the cases of $M=2$ with BICM utilizing both $10$ and $30$ decoding iterations has better energy efficiency than the case of $M=4$, BICM, and $1$ decoding iteration.
Similar behavior is observed in comparing $M=4$ to $M=8$.
This result implies an energy efficiency trade-off - particular error rate and energy operating points may be achieved by varying $M$ or the number of decoding iterations.
The absolute performance improvement of BICM-ID vs. BICM for $M=4$ and $8$ is approximately $0.5$ and $0.7$ dB, respectively, implying that relative performance improvement increases with modulation order.

\begin{figure}[t]
\centering
\includegraphics[width=8.8cm]{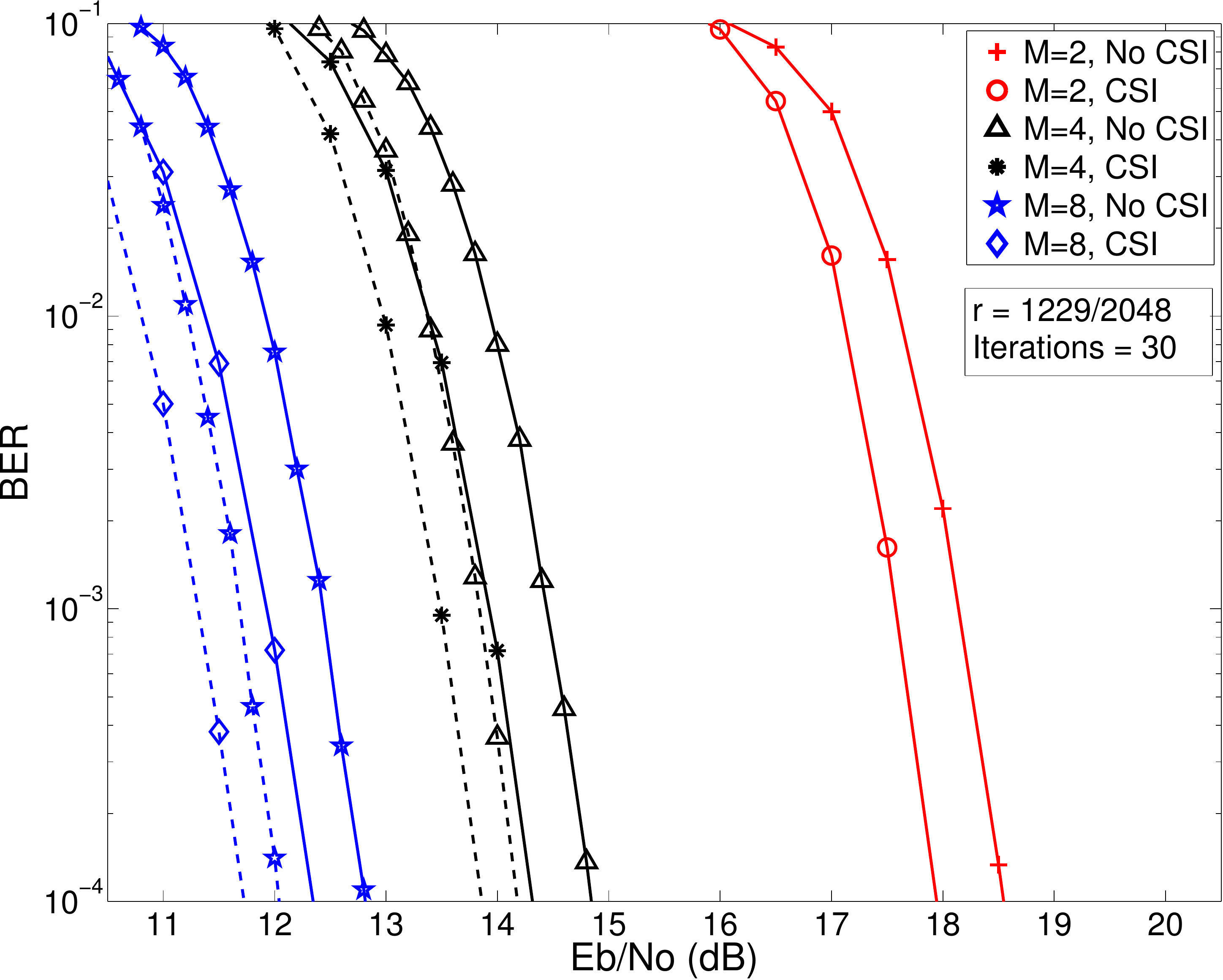}
\caption{Error-rate performance of the relay demodulator as a function of modulation order and channel state information.
The number of decoding iterations for all curves is 30.
The data sequence is length K = 1229.
The channel code rate is $r=1229/2048 \approx 0.6$.
Solid lines denote BICM, and dashed lines denote BICM-ID.
}
\label{fig:2}
\end{figure}

Fig. \ref{fig:2} shows error rate performance of the relay demodulator as a function of modulation order and channel state information.
The number of decoding iterations is fixed at $30$.
The relative performance as a function of CSI is roughly the same between curves having the same modulation order and decoder feedback.
The absolute performance improvement of BICM-ID over BICM considering $M=4$ is about $0.75$ and $0.5$ dB for no CSI, and CSI, respectively.
For $M=8$, the performance improvement is about $0.9$ and $0.6$ dB. 
The implication is that BICM-ID provides a greater relative improvement in the absence of channel state information and with increasing $M$.
A notable result is that BICM-ID with no CSI has better energy efficiency than BICM with CSI.
This implies that decoder complexity may be closely traded off with channel knowledge to achieve a given energy efficiency.

\begin{figure}[t]
\centering
\includegraphics[width=8.8cm]{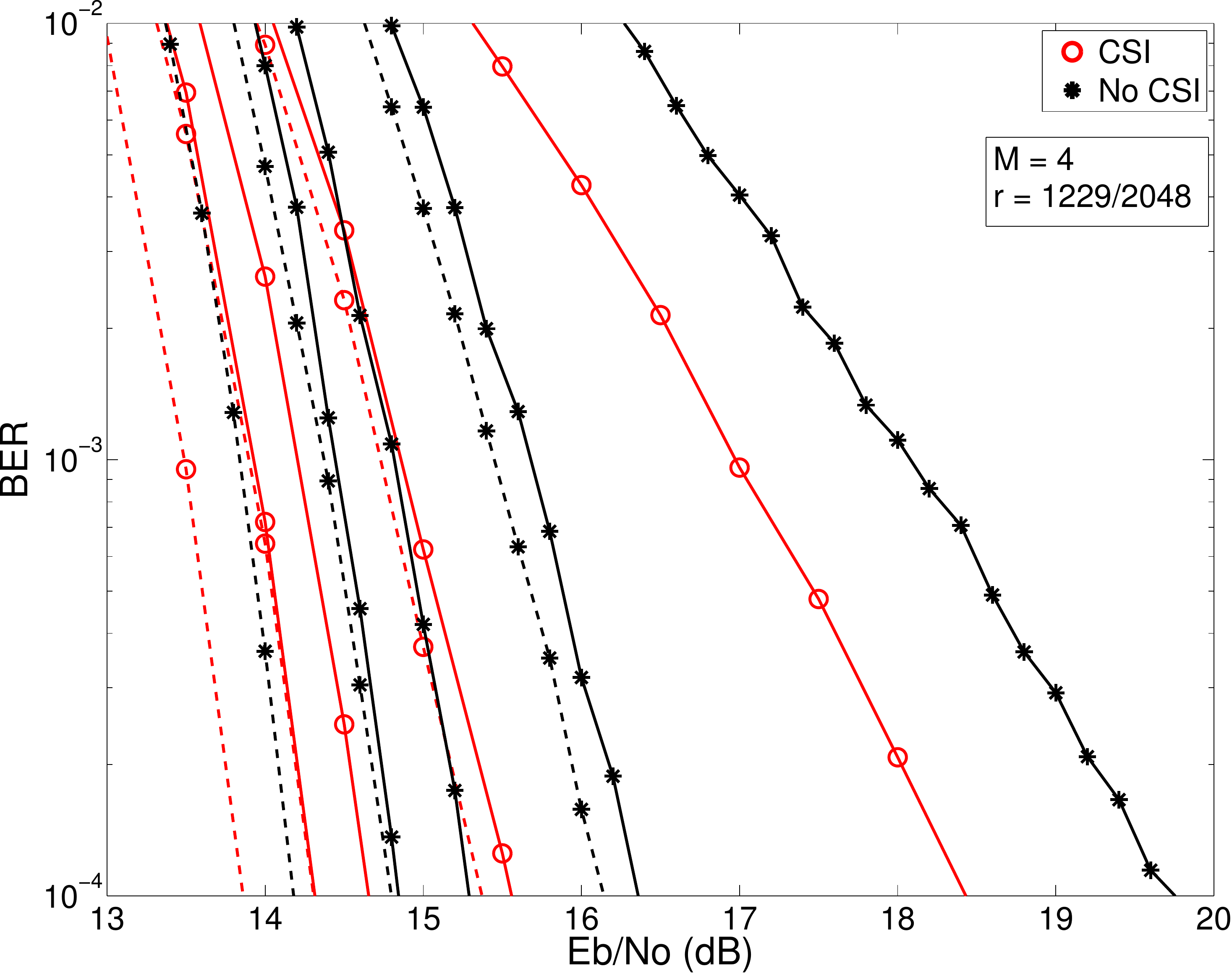}
\caption{Error rate performance of the relay demodulator as a function of decoding iterations and channel state information.
The modulation order is $M=4$.
The data sequence is length K = 1229.
The channel code rate is $r=1229/2048 \approx 0.6$.
Solid lines denote BICM, and dashed lines denote BICM-ID.
Within each case of channel state information, from right to left, the number of decoding iterations is 1,2,4,30.
}
\label{fig:3}
\end{figure}

Error rate performance as a function of decoding iterations and channel state information is shown in Fig. \ref{fig:3}.
The modulation order is fixed at $M=4$.
Considering BICM curves, the performance difference between CSI and no CSI for a given iteration decreases as the number of iterations is increased.
In the case of $1$ iteration, the performance difference is about $2$ dB, while for $30$ iterations, the difference is about $0.5$ dB.
The same property holds for BICM-ID.
BICM-ID allows the no CSI cases to outperform the CSI cases by varying the number of decoding iterations.
Consider the case of no CSI, BICM-ID and $4$ decoding iterations.
This case outperforms the CSI cases using both BICM and BICM-ID using $2$ iterations.
This example clearly illustrates the potential design trade-off in utilizing decoder iterations and channel state information.
Likewise, for $30$ iterations, the no CSI case with BICM-ID outperforms all CSI cases except the case also utilizing $30$ iterations
and BICM-ID.

\subsection{Binary Information Rate}

This subsection presents simulated values of the binary information rate of the relay demodulator as a function of modulation order and channel state information.
Binary information rate is a measure of maximal throughput achievable for a particular BICM receiver configuration and channel model as a function of signal to noise ratio.
The metric can also be interpreted as the minimum energy required for error-free when a capacity approaching code is used.

Binary information rate is computed by \cite{valenti:2005}

\begin{align}\label{eqn:rate}
R = 1 - \frac{\log_2(e)}{\mu} \sum_{k=0}^{\mu-1} E \left[ \max*(0, \Lambda( b_k )(-1)^{b_k} \right]
\end{align}

\noindent where $b_k$ is the $k$-th network-coded bit associated with a particular super-symbol.
and $\Lambda(b_k)$ is the log-likelihood ratio associated with the bit.
The expectation in (\ref{eqn:rate}) is computed by Monte Carlo simulation.

A sequence of bits having length $K=10000$ are generated by each user, divided into groups having
length $\mu$.
Each group is mapped to symbol, modulated, and transmitted over the channel described in Section \ref{sec:sysm}.
The bitwise LLRs of the corresponding super-symbols are computed at the relay using the demodulator
described in Section (\ref{sec:demod}).
The LLRs and network coded bits in the frame are substituted into (\ref{eqn:rate}) to compute
the value of $R$.
The expectation in (\ref{eqn:rate}) is computed over one million trials and for several
values of symbol energy to noise ratio $\mathcal{E}_S/N_0$.
The bit energy to noise ratio can then be computed as $\mathcal{E}_b/N_0 = \mathcal{E}_S/N_0/R$. 

\begin{figure}[t]
\centering
\includegraphics[width=8.8cm]{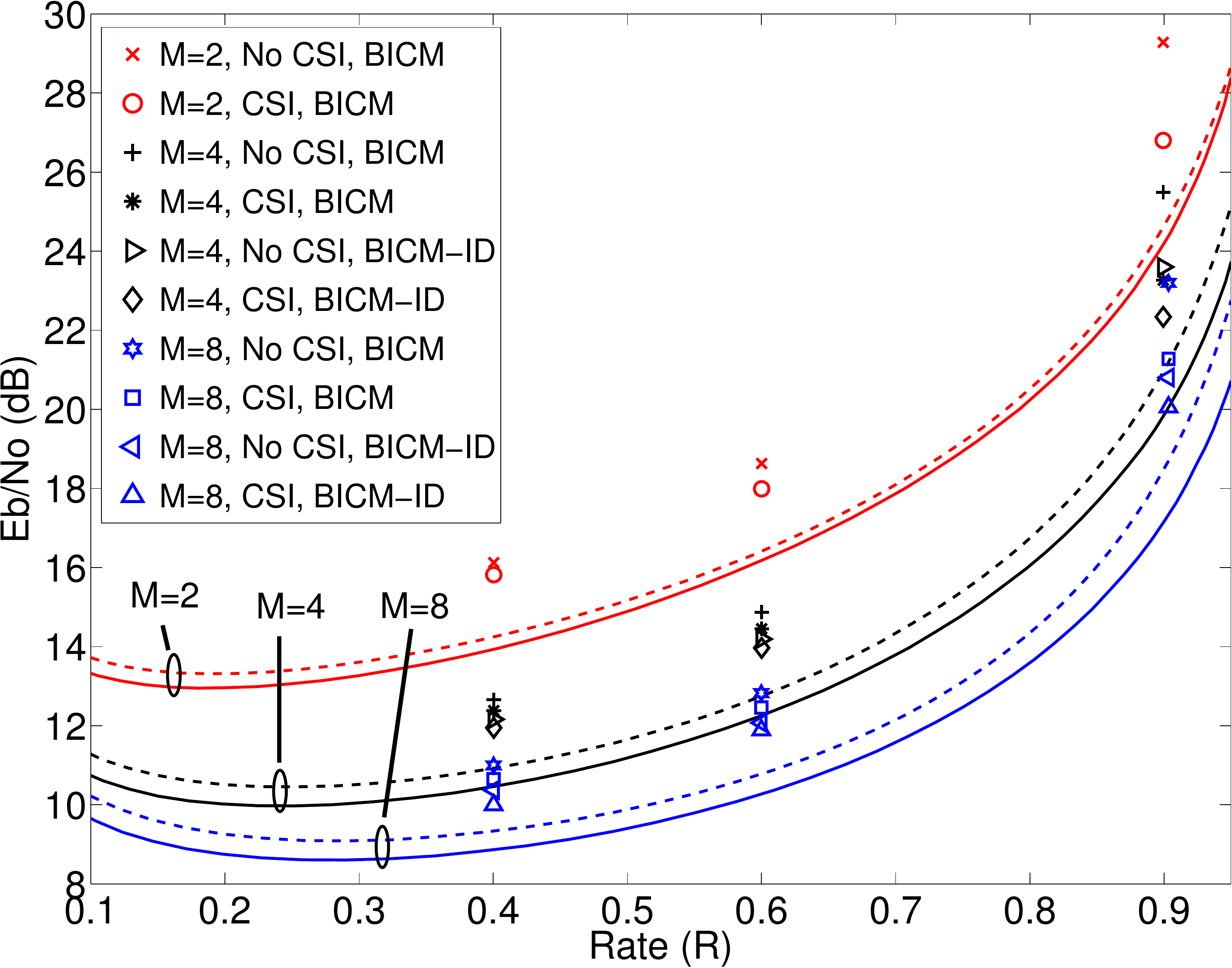}
\caption{BICM capacity at output of relay demodulator as a function of modulation order and channel state information.
Dashed and solid lines denote no CSI and CSI, respectively.
For reference, the $E_b/N_0$ value required to reach an error rate of $10^{-4}$ for several feedback configurations and channel states are shown.
All receivers perform 30 decoding iterations.
}
\label{fig:cap}
\end{figure}

Simulated values of binary information rate are shown in Fig. \ref{fig:cap}.
From top to bottom, the dashed lines and solid lines represent binary information rate with no CSI and CSI respectively, for modulation orders
$M=\{2,4,8\}$.
Energy efficiency improves with increasing modulation order, as expected using orthogonal modulation.
The gap between the energy efficiency with and without CSI increases with modulation order.
A design implication is that channel estimation may be more beneficial in systems utilizing higher modulation orders.
The minimum of each curve represents the most energy-efficient rate for a given level of CSI and modulation order.
For $M=2,\ 4,\ 8$, the most energy efficient rates are approximately $0.2$, $0.26$, and $0.3$, respectively.
The design consideration is that higher throughputs are achievable using less energy as modulation order is increased.

The rate curves shown in Fig. \ref{fig:cap} represent the achievable energy efficiency for a given code rate.
Actual systems will operate at efficiencies which are suboptimal.
Improvements in system design yield improvements in energy efficiency.
In this work, the application of feedback from demodulator to decoder using BICM-ID improves efficiency over BICM
which does not utilize feedback.

Several points representing the $E_b/N_0$ required to reach a simulated error rate of $10^{-4}$ are shown on Fig. \ref{fig:cap}.
For a particular rate and modulation order, points are shown for simulation cases considering different receiver configurations and CSI levels.
For all modulation orders, we see that the efficiency difference between CSI and no CSI receivers increases with code rate.
This implies that channel estimation yields higher performance improvement as code rate increases.
For $M=\{4,8\}$, BICM-ID improves efficiency over BICM, with higher performance improvement as code rate increases.

Considering the trade-off between decoder feedback and channel state information at a particular modulation order, we observe that at code rates $R=0.4$ and $R=0.6$, a system applying BICM-ID with no CSI outperforms a system applying BICM with CSI.
This implies that decoder feedback may be traded off with channel state information as a means of performance improvement at low rates.
At rate $R=0.9$, considering the configuration of BICM with no CSI, applying channel state information yields more performance improvement than BICM-ID.

\balance
\section{Conclusion}
This work presents a relay receiver capable of performing digital network coding in the two-way relay channel
using soft-input channel decoding and feedback from decoder to demodulator, termed BICM-ID.
Simulation results using the UMTS Turbo code, $4$, and $8$-ary modulation, and different levels
of channel state information show error rate improvements between $0.4$-$0.9$ dB over non-BICM-ID systems.
A natural extension of this work is adapting the relay receiver to support coding schemes that perform
channel decoding on the sums of received symbols, rather than network-coded bits, which has been shown to improve
capacity \cite{zhang:2009}.

\bibliographystyle{IEEEtran}
\bibliography{../bib/bibliography}

\end{document}